# Revisiting solving procedure for Ermakov-Pinney equation (with applications in the field of cosmology)

**Sergey Ershkov***, Affiliation: Plekhanov Russian University of Economics,

Scopus number 60030998, 36 Stremyanny Lane, Moscow, 117997, Russia,

Corresponding author, e-mail: sergej-ershkov@yandex.ru,

**Victor Christianto**, Malang Institute of Agriculture, Jalan Soekarno-Hatta Malang, Jawa Timur Indonesia, e-mail: victorchristianto@gmail.com,

**Elbaz I. Abouelmagd**, Celestial Mechanics and Space Dynamics Research Group (CMSDRG), Astronomy Department, National Research Institute of Astronomy and Geophysics (NRIAG), Helwan, Cairo 11421, Egypt, e-mail: elbaz.abouelmagd@nriag.sci.eg

## Abstract

It is known that Ermakov-Pinney equation is a nonlinear equation with wide applications in dynamics, physics, cosmology (e.g., Ermakov equation can be connected to Bose-Einstein Condensate cosmology which unifies the dark energy and the dark matter). In this analytical study, we have presented a new type of solving procedure to obtain analytical solution of Ermakov-Pinney equation, specifically for the case of rotating early Universe with vortex. The particular case of solution of the aforementioned equation is presented also (such the solution of special kind is important for cosmological applications) which corresponds to the class of solutions with symmetry reduction.

**Keywords:** Ermakov-Pinney equation, Bose-Einstein Condensate cosmology, dark energy, dark matter, symmetry reduction.

## Introduction.

Ermakov-Pinney equation [1-2] was introduced more than a century ago, with many applications in various domains of science. This equation is generally connected with the names of Ermakov who presented it in a paper on integrability of second-order ordinary differential conditions and Pinney who, exactly 70 years after the fact of first introducing of such equation, gave the appropriate arrangement (*videlicet,* full clarification) with respect to its general solution in one of the more brief papers in Mathematics [3-4].

There are a lot of useful applications of Ermakov-Pinney equation in various fields of Physics [5-11]. For instance, Tsekov has discussed Ermakov-Pinney equation in Quantum Mechanics [8-9]. It can be shown, if we start with Madelung equations, they will yield the universal form of a dissipative Ermakov equation [9]. Other fields of application include angular momentum and also quantum physics, see for instance [6-7].

Of particular importance is its use in cosmology field. In Hawkins & Lidsey's paper [5], it is stated that it can be shown that the elements of cosmologies sourced by a combination of amazing liquids and self-connecting scalar fields are portrayed by the non-direct, Ermakov-Pinney condition.

While it can be shown that Ermakov equation can be connected to Bose-Einstein Condensate cosmology [10] (which unifies the dark energy and the dark matter)

$$\frac{d^2a}{dt^2} + (\frac{d\phi}{dt})^2 a = \frac{k}{a^3}$$

where $a(t)$ is the scale factor of the universe, $t$ represents cosmic (proper, large scale) time, $k$ is the curvature constant and units are chosen such that Newton's constant $G$ is given by $4\pi G = 1$, and by interpreting the matter source as a self–interacting scalar field, $\varphi = \varphi(t)$ [10].

Nonetheless, thus far only few papers discuss Ermakov-Pinney equation for early rotating Universe with vortex [4]. In an earlier paper [4] the equation, which can be associated with the family of ordinary differential equations of similar type, has been solved numerically using Mathematica computer algebra package. In this paper, the authors will solve Ermakov-Pinney equation for the special case (useful for cosmological applications) in analytical way, which corresponds to the class of solutions with symmetry reduction.

## 1. <u>Equations of the problem.</u>

Despite of being actual for various field of physics (mainly, in cosmology) up to nowdays, Ermakov-Pinney equation [1-2] was introduced first time in the year 1880 [1] as the effective example of novel mathematical approach in solving ODEs of 2-nd order:

$$\frac{d^2 z}{d x^2} = M z + \frac{\alpha}{z^3} , \qquad (1.1)$$

where $\alpha$ = const, but $M = M(x)$ is defined as function which should be determined depending on the solutions of the coupled ODE:

$$\frac{d^2 y}{d x^2} = M y , \qquad (1.2)$$

which can be transformed by simple change of variables $v = (y'/ y)$ to the *Riccati* ODE of 1-st order.

Then hereafter Ermakov obtained first integral [3] (by linear combining along with further integration of Eqns. (1.1)-(1.2))

$$\left( y \frac{d z}{d x} - z \frac{d y}{d x} \right)^2 = C - \alpha \left( \frac{y}{z} \right)^2 , \qquad (1.3)$$

where $C$ = const. It is worthnoting that if we denote $u = (z/y)$ as new variable for the solution of Eqn. (1.2), given by the chosen initial conditions and the appropriate choice of the function $M = M(x)$, we could obtain from (1.3) as below

$$\left(y^2\left(\frac{z}{y}\right)'\right)^2 = C - \alpha\left(\frac{y}{z}\right)^2, \quad \Rightarrow \quad \frac{1}{2}(u^2)' = \pm\frac{\sqrt{Cu^2 - \alpha}}{y^2},$$

$$\Rightarrow \quad \{u^2 = w\} \quad \Rightarrow \quad \int\left(\frac{dw}{\sqrt{Cw - \alpha}}\right) = \pm 2\int\left(\frac{dx}{y^2(x)}\right) \qquad (1.4)$$

here symbol ′ denotes differentiating with respect to argument $x$. So, we obtain from (1.4)

$$\frac{2\sqrt{Cw - \alpha}}{\sqrt{C}} = \pm 2\int\left(\frac{dx}{y^2(x)}\right) \quad \Rightarrow \quad z = y\cdot\sqrt{\frac{\alpha}{C} + \left(\int\left(\frac{dx}{y^2(x)}\right)\right)^2} \qquad (1.5)$$

We will consider furthermore and below in the current research the additional important case (in cosmological applications) when $M$ = const which stems from the class of solutions with symmetry reduction.

## 2. Solving procedure for the Eqns. (1.1)-(1.2) when *M* = const.

Let us consider equation (1.1) in the important case (for the cosmological applications) when $M$ = const, which could be associated with simple classical periodic solutions of equation (1.2)

$$\frac{d^2 z}{d x^2} = M z + \frac{\alpha}{z^3}, \quad \Rightarrow \left\{ \frac{d z}{d x} = p(z) \quad \Rightarrow \quad \frac{d^2 z}{d x^2} = \frac{d\, p(z)}{d z} \cdot p(z) \right\}$$

$$\Rightarrow \frac{d\, p(z)}{d z} \cdot p(z) = M z + \frac{\alpha}{z^3}, \quad \Rightarrow \frac{1}{2} d\left(p^2(z)\right) = \frac{M}{2} d\left(z^2\right) - \frac{\alpha}{2} d\left(\frac{1}{z^2}\right),$$

$$\Rightarrow p^2(z) = M z^2 - \frac{\alpha}{z^2} + C, \quad \{C = const\} \quad \Rightarrow \quad p(z) = \pm \sqrt{M z^2 - \frac{\alpha}{z^2} + C},$$

$$\left\{ \left( M z^2 - \frac{\alpha}{z^2} + C \right) \geq 0 \right\}$$

$$\Rightarrow \quad \int \frac{d z}{\sqrt{M z^2 - \frac{\alpha}{z^2} + C}} = \pm \int d x \quad \Rightarrow \quad \int \frac{d(z^2)}{\sqrt{M z^4 + C \cdot z^2 - \alpha}} = \pm 2 \int d x \qquad (2.1)$$

where

$$\int \frac{d(z^2)}{\sqrt{M z^4 + C \cdot z^2 - \alpha}} \equiv \begin{cases} \frac{1}{\sqrt{M}} \ln \left| 2 M z^2 + C + 2\sqrt{M} \cdot \sqrt{M z^4 + C \cdot z^2 - \alpha} \right| & if\ M > 0,\ C^2 \neq -4M \cdot \alpha \\ \frac{1}{\sqrt{M}} \ln \left| 2 M z^2 + C \right| & if\ M > 0,\ C^2 = -4M \cdot \alpha \\ -\frac{1}{\sqrt{-M}} \arcsin \left( \frac{2 M z^2 + C}{\sqrt{C^2 + 4M \cdot \alpha}} \right) & if\ M < 0,\ C^2 > -4M \cdot \alpha \end{cases} \qquad (2.2)$$

If we choose the appropriate initial data in (2.2), adjusted to the chosen meaning of parameters $\{M, \alpha, C\}$ $(z(x_0) \neq 0)$:

$$\left. \frac{d z}{d x} \right|_{x_0} = \pm \sqrt{M z^2(x_0) - \frac{\alpha}{z^2(x_0)} + C}$$

then we can refer to the 2nd or 3rd branches of solution (2.2), presented above.

In this case, the final expression for $z(x)$ can be presented (after re-inversing procedure with respect to the argument $x$) as below:

$$1)\quad if\ M>0,\quad C^2=-4M\cdot\alpha \quad\Rightarrow\quad \left|2M z^2+C\right| = \exp\left(\pm 2\sqrt{M}\int dx\right)$$

$$if\quad C \geq -2M z^2(x_0) \quad\Rightarrow\quad z = \pm\sqrt{\frac{\exp\left(\pm 2\sqrt{M}\int dx\right)-C}{2M}} \tag{2.3}$$

$$2)\quad if\ M<0,\quad C^2>-4M\cdot\alpha \quad\Rightarrow\quad -\frac{1}{\sqrt{-M}}\arcsin\left(\frac{2M z^2+C}{\sqrt{C^2+4M\cdot\alpha}}\right) = \pm 2\int dx \quad\Rightarrow$$

$$z = \pm\sqrt{\frac{\pm\left(\sqrt{C^2+4M\cdot\alpha}\right)\sin\left(\frac{2}{\sqrt{-M}}\int dx\right)-C}{2M}} \tag{2.4}$$

## 3. <u>Discussion.</u>

As we can see from the derivation above, equations (1.1)-(1.2) are proved to be very hard to solve analytically. Nevertheless, we have succeeded in obtaining analytical formulae for the solution in case which is important for the cosmological applications (*videlicet*, when $M$ = const). Let us clarify that at transforming of equation (1.1) to the form (1.4)-(1.5) by virtue of special change of variables we have taken into account that independent variable (argument $x$) is not included to the left and right parts of (1.1) if $M$ = const (it is worthnoting that invariant (1.3), obtained earlier in [1], has been used during this procedure). So, we have reduced ordinary differential equation of 2-nd order (1.1) by the elegant change of variable

$\left\{ \dfrac{d\,z}{d\,x} = z' \equiv p(z) \;\Rightarrow\; z'' = \dfrac{d\,p}{d\,z}\cdot p \right\}$ to the 1-st order differential equation. Then, having solved equation with regard to function $p(x)$, we should solve ODE in regard to $p = \dfrac{d\,z}{d\,x} = \pm\sqrt{M\,z^2 - \dfrac{\alpha}{z^2} + C}$ to obtain the final result.

It is worthnoting that, according to [3], Pinney simply wrote the solution of (1.1) in [2] as $z(x) = \sqrt{A\,f^{\,2}(x) + 2B\,g(x)\,f(x) + D\,g^{\,2}(x)}$ where the constants, $A$, $B$ and $D$, are related according to $AD - B^2 = \alpha/W$, functions $f(x)$ and $g(x)$ are linearly independent solutions of (1.2) and $W$ is their, obviously nonzero, Wronskian.

## 4. Conclusion.

In this paper, we have presented a new type of the solving procedure to obtain solution of the Ermakov-Pinney equation (1.1) when $M$ = const (stems from the class of solutions with symmetry reduction) which can be associated with the family of ordinary differential equations of similar type corresponding to rotating early Universe with vortex, as it was discussed in [4].

The significance of this paper is to prove that analytical solution of such the type does exist, although with certain assumptions. It can be expected that further investigations for analytical solutions of Ermakov equation can be done.

The last but not least, it is worth noting that as we know, there are a lot of discussions on DM (dark matter) and DE (dark energy) recent years, but no real observations confirming conclusions of these fancy theories. As far as we know, there are three approaches we can consider: (a) dark matter and dark energy originated from certain kind of negative mass physics, we can call it Terletsky-Winterberg negative mass propulsion, (b) possibly also a variation of non-Diophantine arithmetics can be considered for DE and DM origination [12], (c) so-called enthropy lensing may be responsible for distortions of classical Newtonian gravity fields when modelling dynamics of distant objects in Universe. The last conception of enthropy lensing means that various physical effects like tidal phenomena or considerable photo-gravitational field of central forces (at explosion of supernova, for example) may be influencing in real physical processes on dynamics of object under investigation.

The true aim in this case is to distinguish whether DM conception is an artificial (false) science which can not be supported by providing with data of real experiments or it is simply a just mathematical artefact without correspondence to reality/experiments. This seems to be rather philosophical question. But we do believe that if we can model kinematics or dynamics of distant objects in Universe without taking into account conception of DM (dark matter) & DE (dark energy), but using instead the real physical effects, we shall prefer to use a physical model which can be confirmed by the data of real observations.

Also, remarkable works [13-18] should be cited, which concern the problem under consideration. We should make a concluding remark regarding that we do not consider hereby the interaction depending time and space case induced Feshbach resonance in experiments, which described by Ermakov equation connected to Bose-Einstein condensate cosmology which unifies the dark energy and the dark matter for example, dynamics of a bright soliton in Bose-Einstein condensates with time-dependent atomic scattering length in an expulsive parabolic potential [15]. We do not consider also two dimensional case and new exact solutions, for example, quantized quasi-two-dimensional Bose-Einstein condensates with spatially modulated nonlinearity [16], as well as both matter rogue wave in Bose-Einstein condensates with attractive atomic interaction [17], and exact soliton solutions and nonlinear modulation instability in spinor Bose-Einstein condensates [18].

**Declarations**

On behalf of all authors, the corresponding author states that there is no conflict of interest.

The datasets generated during and/or analysed during the current study are available from the corresponding author on reasonable request.

In this research, Dr. Sergey Ershkov is responsible for the general ansatz and the solving procedure, simple algebra manipulations, calculations, results of the article in Sections 2-4 and also is responsible for the search of approximated solutions.

Dr. Victor Christianto and Prof. Elbaz I. Abouelmagd are responsible for theoretical investigations as well as for the deep survey in literature on the problem under consideration. All authors agreed with the results and conclusions of each other in Sections 1-4.

**References**:

[1]. Ermakov V. (1880). *Second order differential equations. Conditions of complete integrability*. Universita Izvestia Kiev, Series III, Issue 9, pp.1-25

(English translation: Harin AO, Redactor: Leach PGL (2008)) Applied Analysis and Discrete Mathematics, Issue 2, pp. 123-145.

[2]. Pinney E. (1950). *The nonlinear differential equation* $y''(x) + p(x)y + cy^{-3} = 0$, Proceedings of the American Mathematical Society, Issue 1, p.681.

[3]. Morris R.M., Leach P.G.L. (2015). *The Ermakov-Pinney Equation: its varied origins and the effects of the introduction of symmetry-breaking functions*. See at: https://arxiv.org/abs/1510.08992

[4]. Christianto V., Smarandache F. (2019). *One note Samba approach to Cosmology*. Prespacetime Journal, Vol. 10, no. 6.

[5]. Hawkins R.M., Lidsey J.E. (2002). *The Ermakov-Pinney equation in scalar field cosmology*. Phys.Rev. **D66**, 023523. DOI: 10.1103/PhysRevD.66.023523

[6]. Kevrekidis P.G., Drossinos Y. (2007). *Nonlinearity from linearity: The Ermakov–Pinney equation revisited*. Mathematics and Computers in Simulation, Vol. 74, 196–202.

[7]. Schuch D. (2008). *Riccati and Ermakov Equations in Time-Dependent and Time-Independent Quantum Systems*. Symmetry, Integrability and Geometry: Methods and Applications SIGMA 4, **043**, 16 pages.

[8]. Tsekov R. (2010). *Ermakov equations in Quantum Mechanics*. arXiv: 1002.2545.

[9]. Tsekov R. (2011). *Quantum Diffusion*. Phys. Scr. **83**, number 035004 [arXiv 1001.1071].

[10]. Lidsey J.E. (2004). *Cosmic Dynamics of Bose-Einstein Condensates*. Class. Quant. Grav. **21**, pp.777-786. DOI: 10.1088/0264-9381/21/4/002

[11]. Christianto V., Smarandache F. (2018). *Solving Numerically Ermakov-type Equation for Newtonian Cosmology Model with Vortex*. Prespacetime Journal, Febr. 2018, p.89.

[12]. Czachor M. (2021). *Non-Newtonian Mathematics Instead of Non-Newtonian Physics: Dark Matter and Dark Energy from a Mismatch of Arithmetics*. Foundation of Science, **26**, pp.75-95.

[13]. Malkin I.A., Man'ko V.I., and Trifonov D.A. (1970). *Coherent States and Transition Probabilities in a Time-Dependent Electromagnetic Field*. Phys. Rev. D **2**, 1371.

[14]. Schuch D. (2018). *Quantum Theory from a Nonlinear Perspective*. Part of the book series: Fundamental Theories of Physics (FTPH, volume **191**), Springer International Publishing AG. https://doi.org/10.1007/978-3-319-65594-9

[15]. Liang Z.X., Zhang Z.D., and Liu W.M. (2005). *Dynamics of a Bright Soliton in Bose-Einstein Condensates with Time-Dependent Atomic Scattering Length in an Expulsive Parabolic Potential*. Phys. Rev. Lett. **94**, 050402.

[16]. Wang D.-Sh., Hu X.-H., Hu J., and Liu W.M. (2010). *Quantized quasi-two-dimensional Bose-Einstein condensates with spatially modulated nonlinearity*. Phys. Rev. A **81**, 025604.

[17]. Wen L., Li L., Li Z.D., Song S.W., Zhang X.F., and Liu W.M. (2011). *Matter rogue wave in Bose-Einstein condensates with attractive atomic interaction*. The European Physical Journal D **64**, pp. 473–478.

[18]. Li L., Li Z., Malomed B.A., Mihalache D., and Liu W.M. (2005). *Exact soliton solutions and nonlinear modulation instability in spinor Bose-Einstein condensates*. Phys. Rev. A **72**, 033611.